\documentclass[%
 reprint,
superscriptaddress,
 amsmath,amssymb,
 aps,
 pre,
]{revtex4-1}

\usepackage{graphicx}
\usepackage{dcolumn}
\usepackage{bm}
\usepackage{xcolor}
\usepackage[bf]{subfigure}
\usepackage{multirow}

\begin{document}

\preprint{APS/123-QED}

\title{Transistors: A Network Science-Based Historical Perspective}

\author{Alexandre Benatti}
\author{Henrique Ferraz de Arruda}
\affiliation{
 S\~ao Carlos Institute of Physics,
University of S\~ao Paulo, PO Box 369,
13560-970, S\~ao Carlos, SP, Brazil 
}

\author{Filipi Nascimento Silva}
\affiliation{Indiana University Network Science Institute, Bloomington, IN, USA.
}

\author{Luciano da Fontoura Costa}
\affiliation{
 S\~ao Carlos Institute of Physics,
University of S\~ao Paulo, PO Box 369,
13560-970, S\~ao Carlos, SP, Brazil 
}


\date{\today}

\begin{abstract}
The development of modern electronics was to a large extent related to the advent and popularization of bipolar junction technology. The present work applies science of science concepts and methodologies in order to develop a relatively systematic, quantitative study of the development of electronics from a bipolar-junction-centered perspective. First, we searched the adopted dataset (Microsoft Academic Graph) for entries related to ``bipolar junction transistor''. Community detection was then applied in order to derive sub-areas, which were tentatively labeled into 10 overall groups. This modular graph was then studied from several perspectives, including topological measurements and time evolution. A number of interesting results are reported, including a good level of thematic coherence within each identified area, as well as the identification of distinct periods along the time evolution including the onset and coming of age of bipolater junction technology and related areas. A particularly surprising result was the verification of stable interrelationship between the identified areas along time.
\end{abstract}

\maketitle


\section{\label{sec1}Introduction}

The influence of science and technology on society and the economy has grown steadily along time. A great deal of this phenomenon relates to the autocatalytic potential of scientific-technological developments, in the sense that new resources tend to increase the opportunity for further applications and developments.

To a good extent, the history of human modernity mirrors the history of electronics, from the experiments with the early transistors, based on germanium crystals to state-of-the-art very large scale integrated circuits. Despite its relatively short span of about 100 years, electronics progressed all the way through varying materials and paradigms.  Then, in the 50's, the transistor became a reality, initiating the digital computer revolution.  From then on, we witnessed the appearance of satellite communication, cell phone technology, the internet, the WWW, deep learning, and among others. Who can predict what the next step will be?

Yet, despite its enormous influence on redefining how humans live and interact, the history of electronics is relatively overlooked, at least at a more general level.  Indeed, even the launching of the transistor received moderate media coverage (e.g.~\cite{riordan1997cristal}). Although there are excellent works on the history of electronics (e.g.~\cite{riordan1997cristal, braun1982revolution, gabrys2013digital, brinkman1997history}), the theme is relatively less explored in a more systematic, quantitative context.

The present work aims at applying recently developed science of science (e.g.~\cite{fortunato2018science}) concepts, and methods as the means to try to understand, from real data, additional aspects and relationships between the events in the history of electronics from the specific perspective of BJT-related topics.  This is interesting for several reasons. First, it provides an interesting case for science of science studies, given its multidisciplinarity, interrelationship between academics and industry~\cite{lievana2010relationship}, and the interplay between scientific articles and patents~\cite{meyer2000does}.  Second, it has the potential for better understanding the chain of events, as well as the interrelationship of areas and concepts, that underlies what is probably among the most important scientific-technological advances in recent human history~\cite{meyer2010can,shibata2010extracting,gazis1979influence}.

The term `bipolar junction transistor' (BJT) was used as a reference from which a network of related works was assembled. As a consequence, all results described and discussed in the present work are limited to this perspective, and then only given the methodology and database entries considered.  Other scientometric researches, focusing on different emphases, methods or databases, are likely to yield complementary and diversified results.

We start this work by presenting an overall perspective of electronics, developed in a chronological manner and centered (but not exclusively) on BJT related topics, so as to provide reference and context for the discussion of the obtained results.  Then, we describe the adopted methodology, which involves modern network science concepts and methods.  In particular, we considered the \emph{Microsoft Academic Graph} (MAG)~\cite{sinha2015overview} database, which incorporates scientific articles as well as patents, etc.

Several interesting results have been obtained.  First, the identified areas were characterized by a good level of coherence among the respectively obtained keywords, which allowed tentative labels to be assigned to each of those areas.  The evolution of the identified areas along time presented several interesting features, including the coming of age of several of the involved areas.  When observed in terms of networks, the evolution of the identified areas presented striking conservation of interrelationships, in the sense that the patterns of interconnectivity between the main areas tended to remain nearly the same.

The rest of this paper is organized as follows. We start by presenting a summary panorama of electronics, in Section~\ref{sec:panorama}, aimed at providing some context for readers from other areas. In section~\ref{sec:network}, we describe the methodology employed to obtain the citation network, as well as the adopted network measurements. Section~\ref{sec:res}, presents the obtained results. The conclusions and perspectives of future works are addressed in Section~\ref{sec:conc}.

\section{An Overall Panorama of Electronics}
\label{sec:panorama}
One first important consideration when approaching the history of electronics concerns its relationship and distinction from the history of electricity. Generally speaking, electricity deals with (approximately) linear systems such as power sources and passive devices such as resistors, capacitors, and inductors. In this sense, the main differentiating aspect of electronics regards two points: (i) active devices, capable of amplifying the power of signals (e.g.~transistors); and (ii) non-linearities found in most non-linear devices. Another important aspect is that electronics can be subdivided into major areas, including but not being limited to: (a) linear; (b) digital; (c) power; and (d) high frequency.

To a good extent, the history of electronics follows the development of a sequence of device technologies capable of rectification and, more importantly, power amplification.   One of the first electronic devices was invented by K. F. Braun in 1874, named \emph{crystal detector}~\cite{kurylo1981ferdinand}. This device consisted of a conductor, the \emph{cat whisker} made to touch the surface of a small galena rock. After substantial efforts, this device was capable of unstable electronic detection of rectification (i.e.~directional current conduction), being used in the first rudimentary radios.  Another primitive rectifier was E. Branly's \emph{coherer}, invented in 1890~\cite{dilhac2009edouard}.  This device, basically of a tube filled with iron powder, exhibited a rather limited ability for signal rectification.

An important subsequent event was the discovery of the diode vacuum tube, yet another rectifying device, by J. A. Fleming in 1904~\cite{garson2015birth}.  Basically, this device consists of an incandescent lamp including an additional electrode (the plate, which acts as an anode).  The filament corresponds to the cathode. When positive potential (concerning the cathode) is applied to the plate, the electrons being emitted by the cathode are attracted to it, establishing a respective current.  However, no current is observed when the plate is biased negatively concerning the filament. The vacuum diode allowed substantial improvements in electronic rectification, constituting perhaps the first mark in the history of electronics.

Motivated by continuing demands of long-distance telephony expansion, research efforts were invested in trying to develop a practical device capable of reinforcing the audio signal along the telephonic lines.  The triode, invented by L. de Forest in 1906 under the name of \emph{audion}~\cite{de1906audion} was the first practical device capable of effective amplification of the power of signals, being vastly applied not only in telephony, but paving the way for the unfolding of electronics into a variety of areas including but not being limited to telecommunications (the radio, and then television), space research, medical instrumentation, military (especially radar) and \emph{electronic computing}.  As a consequence of the triode limited frequency response, more sophisticated vacuum tubes including more electrodes were developed.

The early crystal detectors were improved as semiconductor point-contact devices for military applications during World War II, especially at Bell Labs, MIT, and Purdue~\cite{hempstead2005encyclopedia}.  Advances in quantum mechanics at the time finally allowed some understanding of the rectifying action of the crystal diode.  

At about this same time, similar demands, allied to the never-ceasing requirements from telephony and other electronics applications, motivated research aimed at obtaining smaller and more energy-efficient amplifying devices based on semiconductor technology.  The first \emph{transistor} was implemented and demonstrated at Bell Labs by J. Bardeen, W. Brattain, and W. Shockley, in 1947, a substantially important event that received moderate media coverage~\cite{riordan1997cristal}.  These first devices were based on point-contact technology, and further developments were restricted by surface-charge phenomena. Ultimately, alloy devices were obtained, followed by bipolar junction transistors, which became the reference for several decades. 

After World War II, and also during it, great efforts were invested in achieving transistors capable of higher frequency operation, for applications in telecommunications. It also allowed the mass production of very small radio sets. Also at that time, transistors started being used as the basic element, jointly with diodes, in digital computing mainframes, replacing the previous vacuum-tube based machines with an impressive economy of space and energy, while also increasing reliability~\cite{hempstead2005encyclopedia}.

Though semiconductor electronics initially relied strongly on Germanium, it gradually shifted to Silicon, which allowed improve thermal stability.  Several new transistor technologies followed, including FET, JFET, NMOS, CMOS, etc. These developments ultimately led to the concept of integrated circuits, especially through efforts of J. Kilby and R. Noyce, which would soon provide the basis for the personal computer revolution along with the 70's, 80's and 90's.  Also impressive was the development of microcontrollers, which found immediate and wide applications in consumer electronics, among other areas.  

In the analog world, the concept of the solid-state operational amplifier was progressively perfected roughly along this same period, often as a valuable resource in the design and application of frequency and phase filters.  Also worth noticing was the development of hybrid solutions to analog problems, in which digital and analog approaches are integrated, placing a special demand on analog-to-digital and digital-to-analog converters, required to provide the interface with an arguably analog real-world.  One particularly critical aspect here, as the resolution of these converters increases, concerns controlling electromagnetic noise.

Another aspect that has proven to be determinant in several electronics applications concerns the control of electromagnetic interference (EMI), both from the environment into the circuit as well as vice-versa. Another typical concern regards the effect of radiation (e.g. cosmic rays) on semiconductors, such as the possibility of memory errors. Also important is the robustness of semiconductors to electric discharges induced by external charge accumulation. Yet another concern in electronic circuits, especially in power electronics, regards the need to control the operating temperature of the devices. Each of these problems has motivated its related research area and set of respective approaches.

\section{\label{sec:network} Methodology and the employed dataset}
In this section, we describe the employed dataset and the methodology applied to create the network. Furthermore, we present network measurements used in our analyses.  

\subsection{\label{sec:citation}Citation network}
Many studies have been taking into account the organization of papers in terms of citation networks~\cite{fortunato2018science,de2018integrated,silva2016using}. Here, we also consider this type of approach in order to obtain data regarding the transistor's network. The employed dataset is the \emph{Microsoft Academic Graph} (MAG)~\cite{sinha2015overview}, which contains data of millions of research documents, including journal and conference papers, patents, books, and not assigned documents. Our dataset contains data produced between 1926 and 2018. The complete information of the dataset can be found in MAG's web page~\footnote{ {https://www.microsoft.com/en-us/research/project/microsoft-academic-graph/}}.

\begin{figure}[!htpb]
  \centering
     \includegraphics[width=0.99\linewidth]{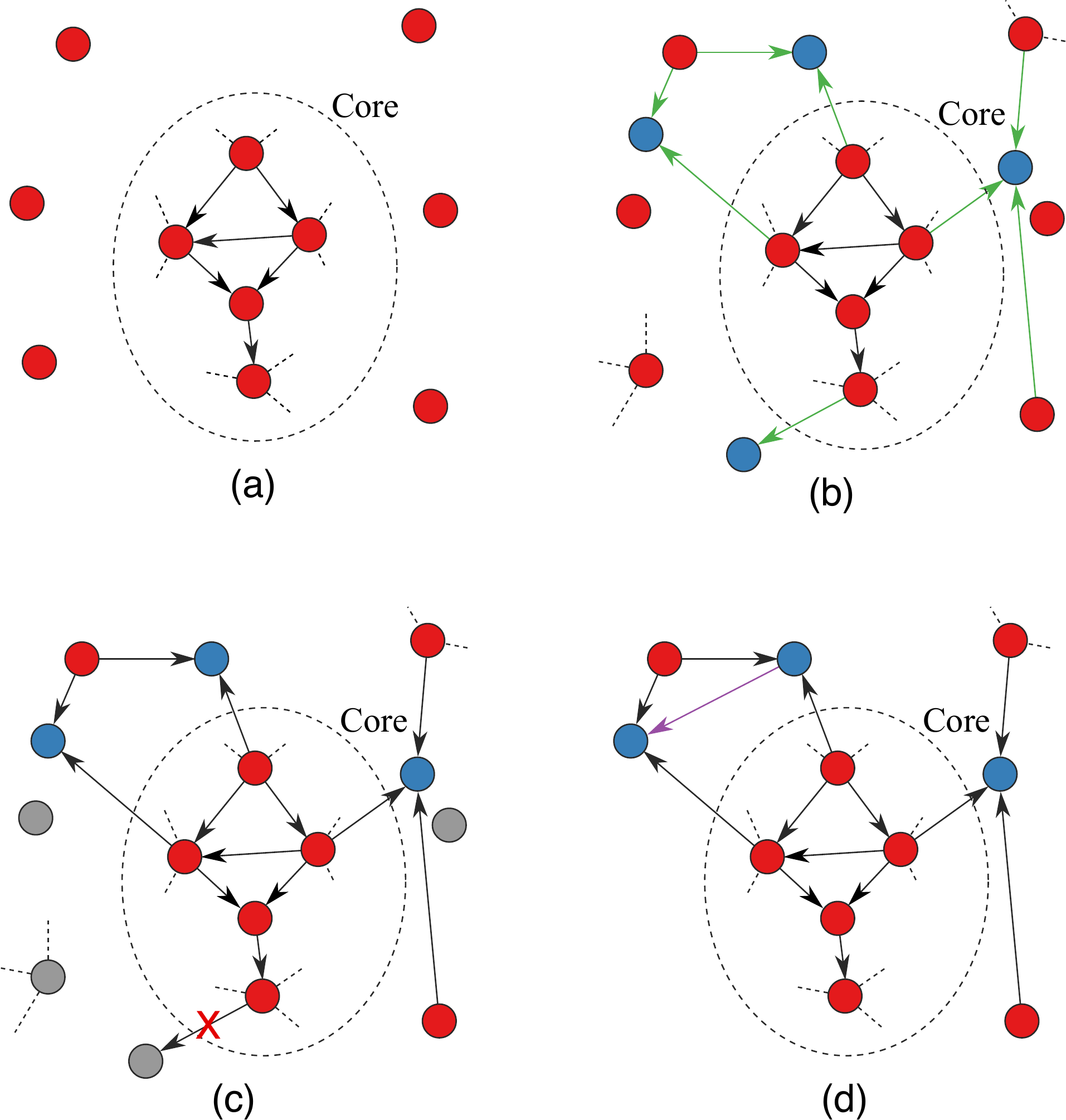}
   \caption{Diagram representing how the studied network is created. The region denoted by the dashed line represents the network core.  (a): the selected nodes (in red), obtained from the keyword search ("bjt" and "bipolar junction transistor") and the respective connections. (b): the cited documents are included own in blue).  (c): blue leaf nodes leading to entries not selected in the search are removed, and the larger connected component is taken. (d): the edges between blued nodes are included.}
  \label{fig:scheme}
\end{figure}

First, by considering the abstracts, we select an initial set of documents with the words \emph{bjt} and \emph{bipolar junction transistor}. In order to obtain a respective network, each document is understood to represent a node, and the edges denote the citations between the documents. First, this network is created, and the largest connected component is identified, which is henceforth called \emph{core} (see Figure~\ref{fig:scheme}~(a)). In order to complement the connectivity between nodes, we also consider the documents cited from the references listed in the initial set, green edges as illustrated in Figure~\ref{fig:scheme}~(b). In the next step, we remove among the blue nodes, those disconnected or with indegree one (leaf nodes), and we consider only the larger connected component as result, as shown in Figure~\ref{fig:scheme}~(c). Finally, we insert the edges between the blue nodes (purple edge), in Figure~\ref{fig:scheme}~(d)), if there is a citation between them. It is important to recall that the direction of the network edges is from the document that cites to those cited.

\begin{figure*}[!htbp]
  \begin{center}
  \begin{subfigure}[Network]{
   \includegraphics[width=0.65\linewidth]{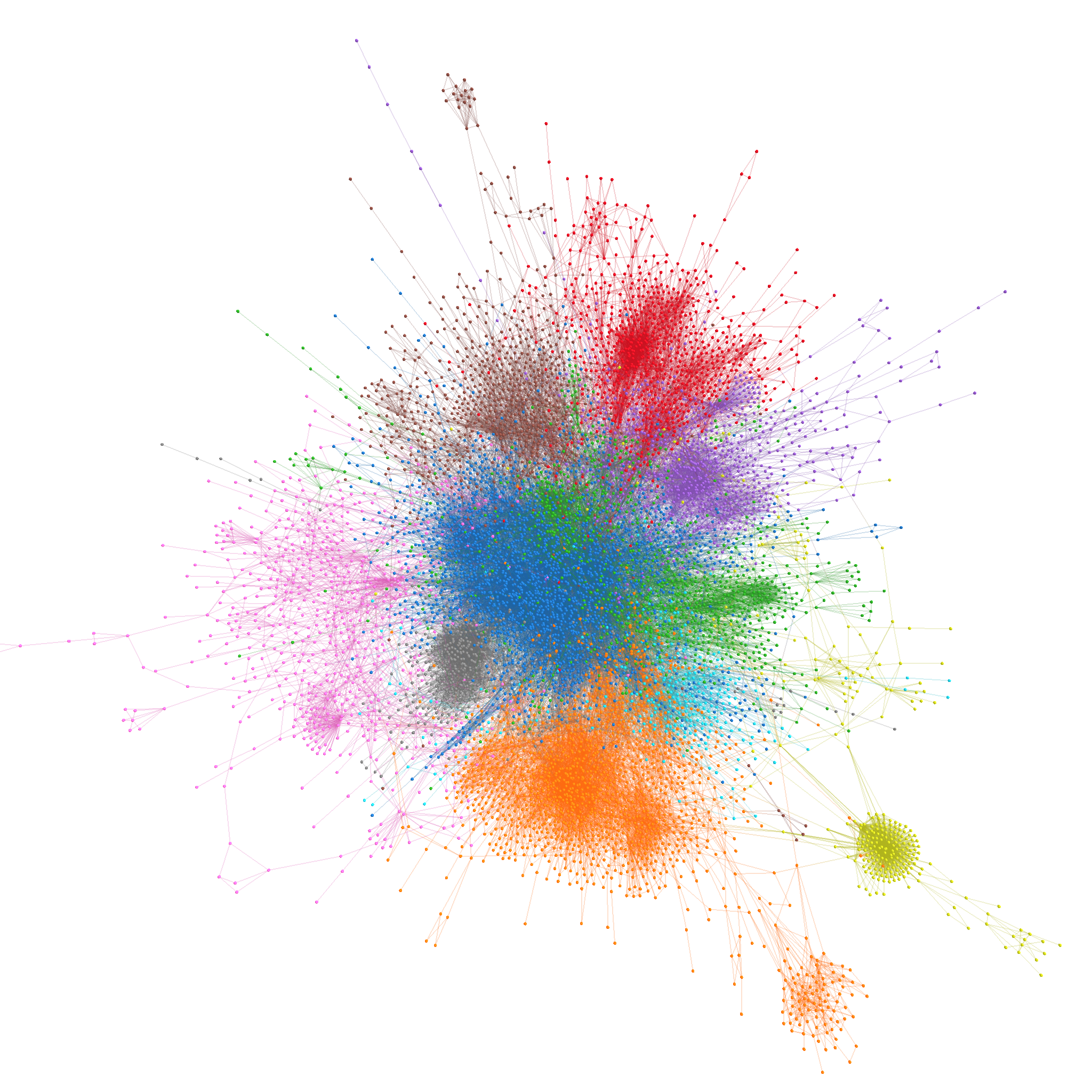}}
  \end{subfigure}
  \begin{subfigure}[Reduced Network]{
   \includegraphics[width=0.65\linewidth]{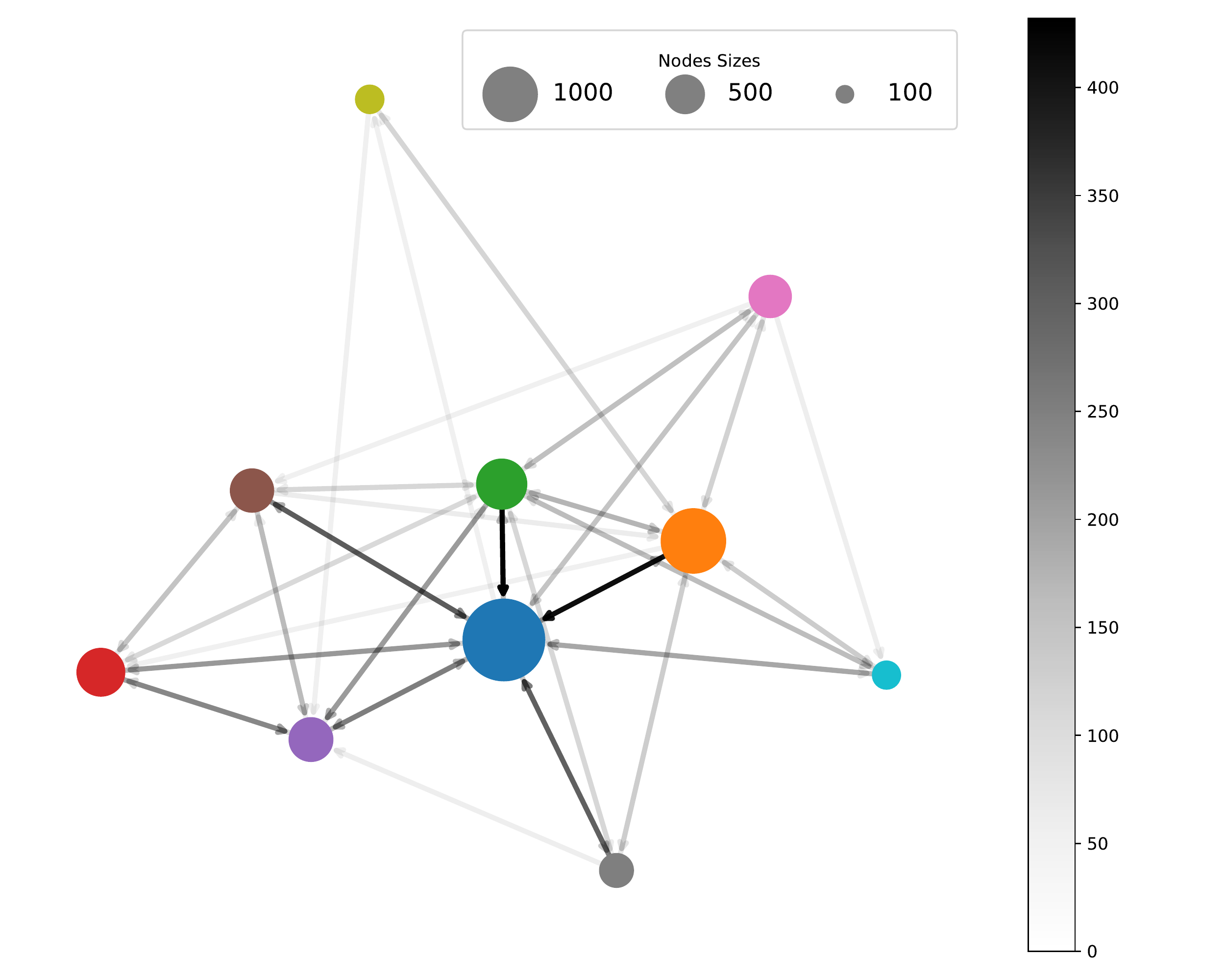}}
  \end{subfigure}
   \includegraphics[width=0.22\linewidth]{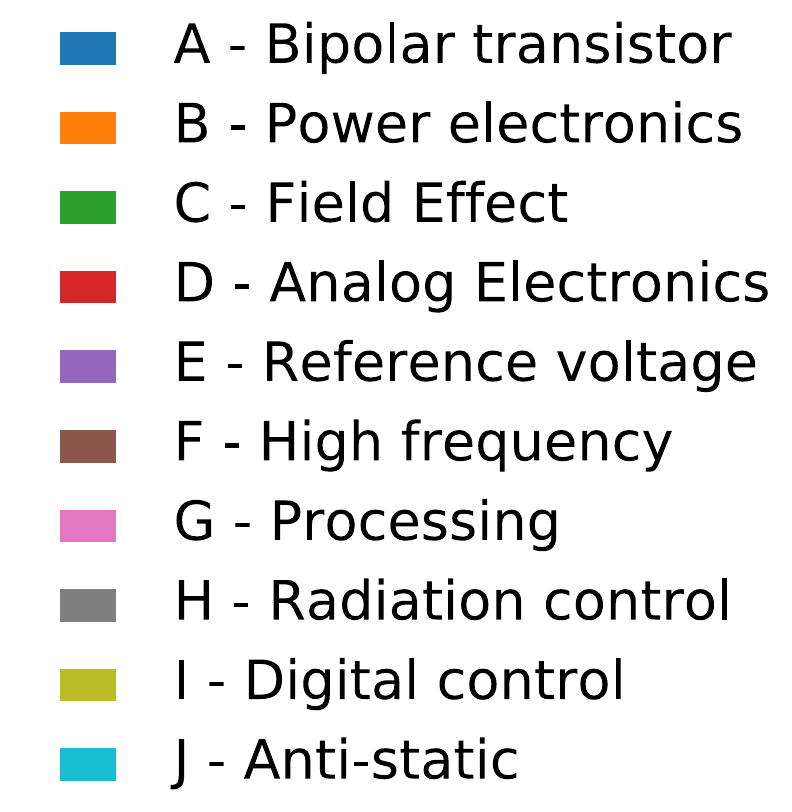}
  \caption{(a) Network visualization, where nodes and edges represent document citations, respectively.  (b) The reduced version of the network. More specifically, the nodes account for the communities, and the edges are weighted according to the number of citations between communities (edges with less than 7 citations are not shown). For both the visualizations, the node positions were calculated from a force-directed algorithm described in~\cite{silva2016using}. \label{fig:network}}
  \end{center}
\end{figure*}

We employed a keyword extraction method based on the community structure of the network, as described in \cite{silva2016using} to derive its main topics. This method starts by detecting the network communities, which are defined as groups of nodes more connected between themselves than to others in the network~\cite{newman2004detecting}. Here, we considered the same methodology variation proposed in~\cite{ceribeli2019coupled}, in which \emph{Infomap}~\cite{rosvall2009map} is employed to detect the community structure of the network, a technique that has been used in related applications of the science of science studies~\cite{rosvall2008maps,rosvall2010mapping}. A visualization of the obtained citation network is shown in Figure~\ref{fig:network}.

\subsection{Network measurements}
In this section we present the network measurements employed in the characterization of the obtained citation network. Because our network is directed, we considered the directed version of degree~\cite{costa2007characterization}. More specifically, degree, $k$, is defined for each node as the number of edges connected to it. In this case, two definitions can be considered, which refers to the counts of in and out edges. In terms of the citation networks, the measurements of in and out degrees represent the number of times a document is cited and cite others, respectively.

Another considered measurement is clustering coefficient~\citep{watts1998collective}. For the sake of simplicity, we considered the undirected version of this measurement, which is defined as follows
\begin{equation}
    c_i = \frac{\lambda(i)}{\tau(i)},
\end{equation}
where $\lambda(i)$ and $\tau(i)$ are the number of triangles of edges and triple connected, respectively. In the citation network, the high clustering coefficient is found for documents that cite or are cited by others that also have edges between them.  

Another measurement that is extensively used is the betweenness centrality~\cite{freeman1977set}, which is calculated for each network node and can be computed as follows
\begin{equation}
B_{k} = \sum_{ij} \frac{\sigma^{k}_{ij}}{\sigma_{ij}} ,
\end{equation}
where $\sigma_{ij}$ is the number of shortest paths that connects $i$ and $j$, $\sigma^{k}_{ij}$ is the number of shortest paths connecting the nodes $i$ and $j$ that crosses $k$. Here, we consider directed paths. This measurement has been employed in studies concerning the analysis of citation networks~\citep{leydesdorff2007betweenness,yan2009applying,lee2008identify}.

\section{\label{sec:res} Results and discussion}
Table~\ref{tab:areas} lists the keywords obtained for each of the 10 identified thematic areas, as well as their respective number of nodes and average degree.  The table is organized in decreasing order of the number of nodes. It should be observed that the labeling of each of these areas is only approximate as an attempt to express the predominant trend among the respectively obtained keywords.  Therefore, the labels of the identified areas are mostly tentative abbreviations of the respective topics.

\begin{table*}[!htbp]
\begin{tabular}{|c|c|c|c|c|}
\hline
Detected key-words & Tentative label & No. of Nodes & Core (\%) & $\langle k \rangle$ \\ \hline
bipolar transistor, model, emitter, heterojunction bipolar, base  & A - Bipolar transistor& 2270 (28.09\%) & 16.70 & 10.97\\
power, h sic, high, h-sic, switch & B - Power electronics  & 1407 (17,41\%) & 25.66  & 9.83 \\
soi, device, gate, lateral, channel & C - Field Effect & 848 (10.49\%) & 21.93 & 8.86 \\
circuit, filter, linear, analog, nonlinear & D - Analog Electronics & 769 (9.51\%) & 12.74  & 9.74 \\
reference, cmos, temperature, voltage, supply & E - Reference voltage & 644 (7.971\%) & 10.87 & 10.07 \\
noise, ghz, frequency, amplifier, db & F - High frequency & 629 (7.78\%) & 13.99 & 6.95 \\
form, region, layer, method, emitter & G - Processing & 601 (7.44\%) & 16.64 & 5.08 \\
radiation, irradiation, dose, damage, effect & H - Radiation control & 382 (4.73\%) & 30.89 & 15.95 \\
control, power, switch, converter, output & I - Digital control & 269 (3.33\%) & 7.06 & 11.08 \\
esd protection, electrostatic discharge, parasitic, device, discharge esd & J - Anti-statics & 263 (3.25\%) & 6.84 & 7.42 \\ \hline
\end{tabular}
\caption{Properties of the identified areas. The keywords, shown in the first column, were automatically obtained, and were then employed to chose the labels of the second column. The average degrees were calculated by considering the undirected version of the network. \label{tab:areas}}
\end{table*}

As could be expected, the major identified areas correspond more closely to BJT theory and modeling, which was considered as the focus of our research. $28.09\%$ of its nodes are part of the core, which is also the largest percentage observed among all obtained areas.  Interestingly, a substantial number of works were incorporated into this area, complementing the original core. The size of the other areas decreases progressively, reflecting not only their intrinsic size but also how their developments relate to BJTs. We observe expressive sizes for power electronics, field effect, and analog electronics. Another interesting result in this table is the diversity of average degree respective to each area, varying from $7.42$ (for anti-statics) to $15.95$ (for radiation control). These average degree variations may indicate that the respective areas have specific levels of integration among its parts, as well as with other areas. For instance, area H seems to be characterized by higher interconnection as far as its average degree is concerned, while area C has about half value for this measurement.

Table~\ref{tab:type} indicates the percentage of types of documents --- patent, journal, book, conference or other --- obtained for each of the 9 identified groups.  Interestingly, a substantial diversity can be observed regarding these measurements, with areas G and I incorporating many patents, while area F presents a relatively high number of conference entries.  Area H is mostly covered in journals.  These results provide further indication of possible heterogeneity between the specific ways in which each identified area has developed and been organized.

\begin{table*}[!htbp]
\begin{tabular}{|c|c|c|c|c|c|}
\hline
\multirow{2}{*}{\textbf{Group Name}} & \multicolumn{5}{c|}{\textbf{Proportion of each document type (\%)}} \\ \cline{2-6} 
 & \textbf{Patent} & \textbf{Journal} & \textbf{Book} & \textbf{Conference} & \textbf{None}  \\ \hline
A - Bipolar transistor & 1.37 & 67.97 & 0.62 & 23.04 & 7.00 \\
B - Power electronics & 2.99 & 59.28 & 0.71 & 30.70 & 6.33 \\
C - Field Effect & 8.96 & 60.61 & 0.83 & 24.53 & 5.07 \\
D - Analog Electronics & 1.43 & 59.69 & 1.69 & 22.50 & 14.69 \\
E - Reference voltage & 19.10 & 38.82 & 0.78 & 33.23 & 8.07 \\
F - High frequency & 2.23 & 43.72 & 1.43 & 43.72 & 8.90 \\
G - Processing & 84.69 & 6.32 & 0 & 8.65 & 0.33 \\
H - Radiation control & 0.26 & 77.75 & 1.05 & 13.09 & 7.85 \\
I - Digital control & 62.83 & 18.22 & 0.74 & 15.61 & 2.60 \\
J - Anti-statics & 14.83 & 36.88 & 1.52 & 41.44 & 5.32 \\
\hline
\end{tabular}
\caption{Proportions of types of documents in each of the identified areas, where \emph{None} represents the uncategorized documents. \label{tab:type}}
\end{table*}

Figure~\ref{fig:network} depicts the overall network obtained from our specific approach, focusing on BJT and using the described data. One first interesting result is the diversity of interconnections observed for each of the identified areas. For instance, the area I is highly compact, while other areas such as D and G are more distributed. As could be expected, the BJT-related area A resulted as the most central area in the obtained network, presenting strong interfaces with many of the other obtained areas.  

Areas G and I, as indicated in Table~\ref{tab:type}, mostly involve patents. In particular, the densely interconnected area I is mostly related to 6 patents~\cite{agarwal2016resistance,melanson2016compensating,maru2016two,zanbaghi2016single,zanbaghi2017detection,melanson2017switch} that are intensively cited by other documents in this identified area. 
Areas B, C, and D present an interesting diversity of interconnections, suggesting the existence of respective communities or modules.

Figure~\ref{fig:network}(b) shows the network obtained by subsuming the nodes in each of the identified areas as a single node.  This representation provides a summarized representation of the interconnection between the identified areas. The centrality of area A is again observed, also receiving more connections than those that are sent to other areas, which is somehow surprising given the way in which the networks was built (see Section~\ref{sec:citation}).

\begin{figure}[!htbp]
\centering
\begin{subfigure}[In-degree]{
\includegraphics[width=0.449\linewidth]{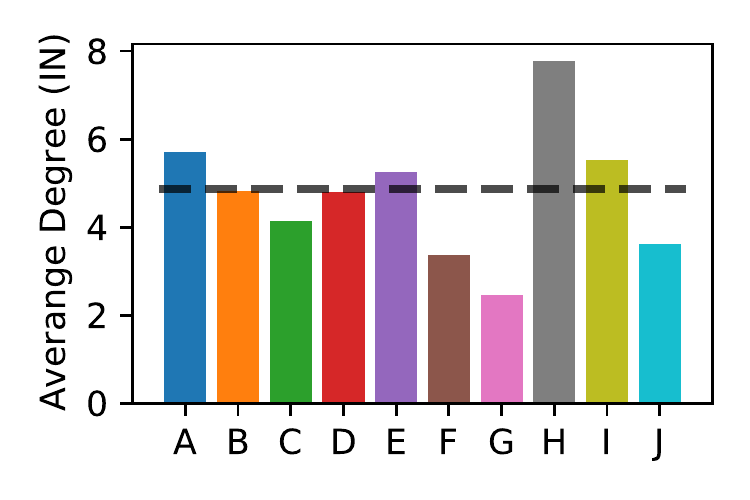}}
\end{subfigure}
~
\begin{subfigure}[Out-degree]{
\includegraphics[width=0.449\linewidth]{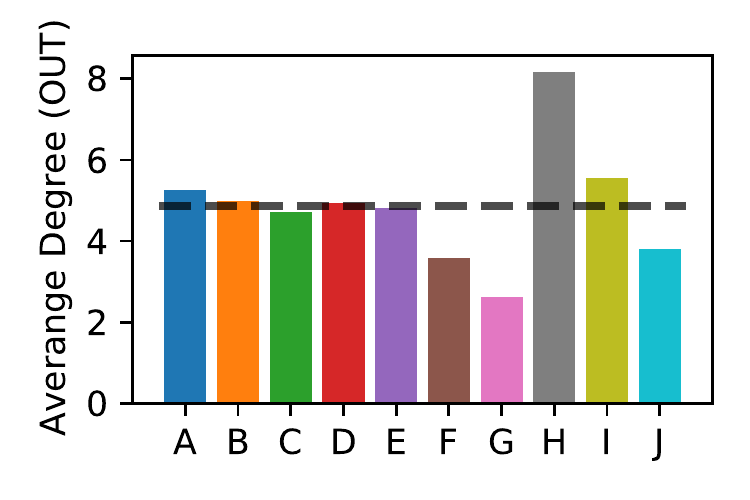}}
\end{subfigure}
~
\begin{subfigure}[Betweenness centrality]{
\includegraphics[width=0.449\linewidth]{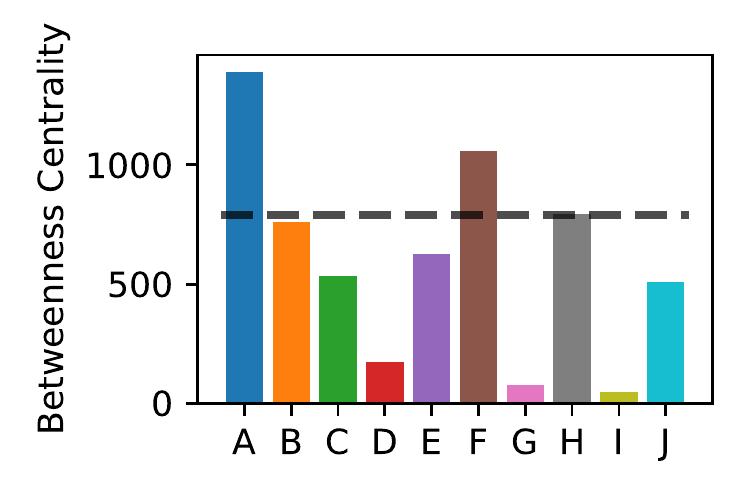}}
\end{subfigure}
 ~
 \begin{subfigure}[Clustering coefficient]{
 \includegraphics[width=0.449\linewidth]{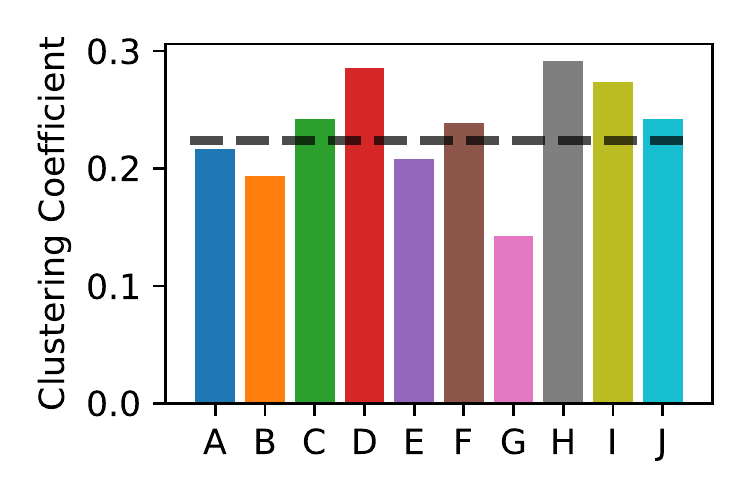}}
 \end{subfigure}

\caption{Each subfigure represents a network measurement calculated for the entire network, in which bars denote the average values of each community.}
\label{fig:hist}
\end{figure}

Figure~\ref{fig:hist} presents the average of the measurements obtained for all nodes of each of the identified areas, but considering them integrated into the overall network.  The in-degree (Figure~\ref{fig:hist}~(a)) and out-degree (Figure~\ref{fig:hist}~(b)) tend to be relatively homogeneous, except for area H presenting larger in- and out-degree, and areas F, G, and J with small degrees.  Also, the in-degrees distribution in (Figure~\ref{fig:hist}~(a)) tends to be very similar to the out-degrees shown in (b). The betweenness centrality, which can be understood as a measurement of the information flow in the network, is shown in Figure~\ref{fig:hist}~(c). As expected, area A was found to be most central. Although F and H are relatively small communities, they have high values of Betweenness centrality, perhaps reflecting their roles in telecommunications, one important application of electronics that underwent an impressive increase along the last decades. Figure~\ref{fig:hist}~(d) shows the clustering coefficient values obtained for each of the areas, which resulted to be relatively small and uniform, indicating similar patterns for the most of the obtained communities.

\begin{figure}[!htbp]
\centering
\begin{subfigure}[\ Degree of groups (in)]{
\includegraphics[width=0.449\linewidth]{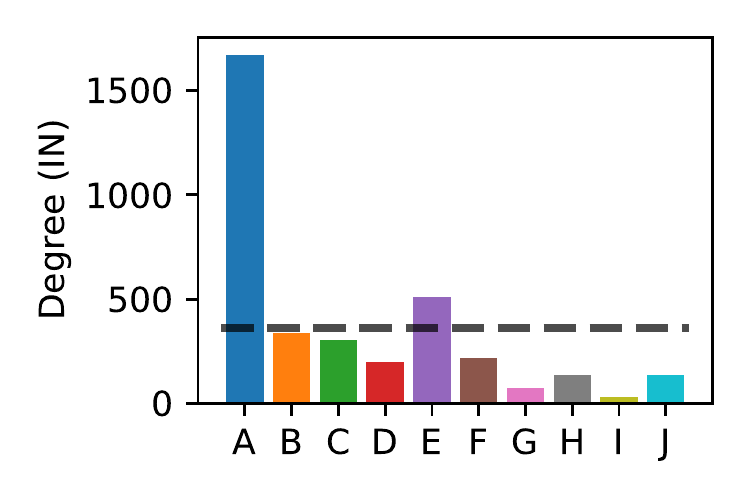}}
\end{subfigure}
\begin{subfigure}[\ Degree of groups (out)]{
\includegraphics[width=0.449\linewidth]{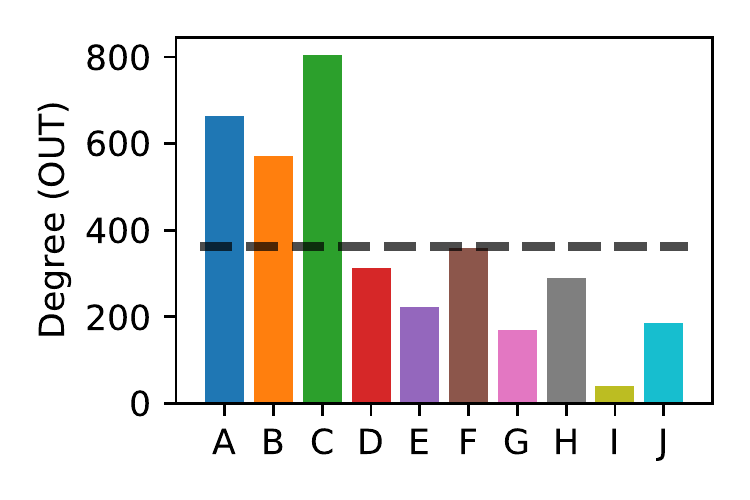}}
\end{subfigure}
\caption{Each subfigure represents a measurement calculated among the obtained communities, in which bars denote the average values of each community.}
\label{fig:hist2}
\end{figure}

Figure~\ref{fig:hist2} depicts the relationship between the obtained communities.  Figure~\ref{fig:hist2}(a) represents the number of citations that a given community receives from the others.  As a complement, we compute out-degree (see Figure~\ref{fig:hist2}(b)), which accounts for the number of times a given community cite others. We observe a high degree of asymmetry between the connections, with area A receiving many more citations than citing the other areas.

\begin{figure*}[!htpb]
  \centering
     \includegraphics[width=0.99\linewidth]{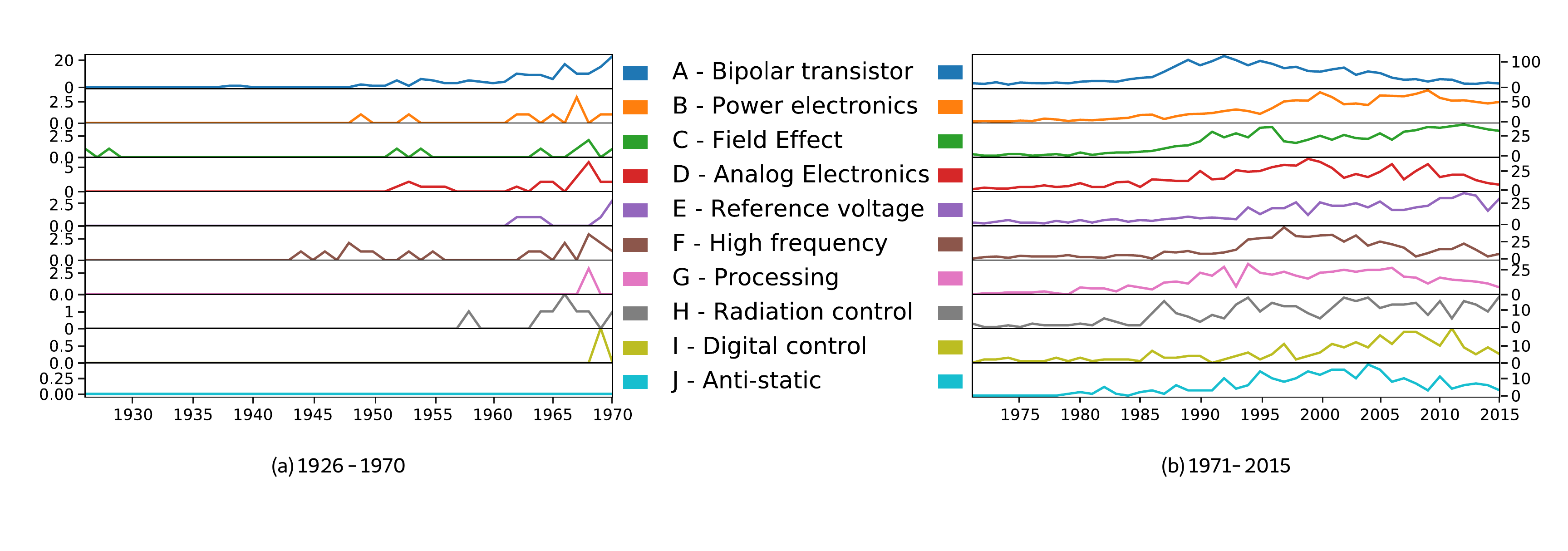}
   \caption{Temporal evolution of papers published along subsequent years for each of the identified areas. Note that we divide the temporal series into two plots given their different scales.}
  \label{fig:time_line}
\end{figure*}

It is also interesting to consider the changes in the number of works in each identified area along time, which is shown in Figures~\ref{fig:time_line}(a-b). For simplicity's sake, we broke these time series according to two subsequent periods: from 1926 to 1970 and from 1971 to 2015. Observe the different scales of the y-axes in these two figures.

\begin{figure*}[!htbp]
\centering
\begin{subfigure}[1950 ($\langle k \rangle = 0.00$ and $n=14$)]{
\includegraphics[width=0.31\linewidth]{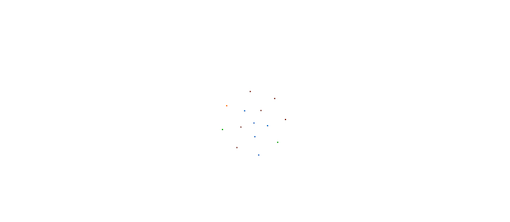}}
\end{subfigure}
\begin{subfigure}[1960 ($\langle k \rangle = 2.00$ and $n=62$)]{
\includegraphics[width=0.31\linewidth]{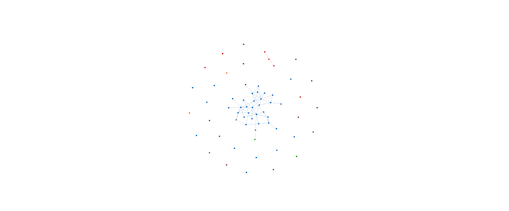}}
\end{subfigure}
\begin{subfigure}[1970 ($\langle k \rangle = 2.92$ and $n=234$)]{
\includegraphics[width=0.31\linewidth]{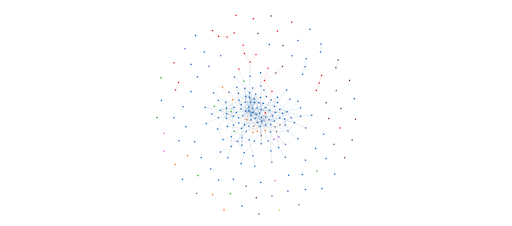}}
\end{subfigure}

\begin{subfigure}[1980 ($\langle k \rangle = 3.21$ and $n=570$)]{
\includegraphics[width=0.31\linewidth]{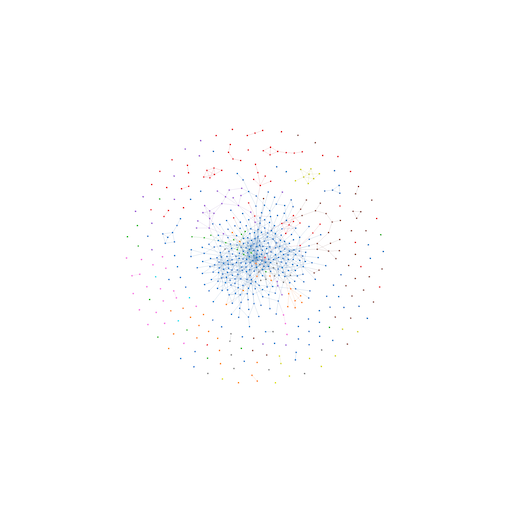}}
\end{subfigure}
\begin{subfigure}[1990 ($\langle k \rangle = 4.68$ and $n=1682$))]{
\includegraphics[width=0.31\linewidth]{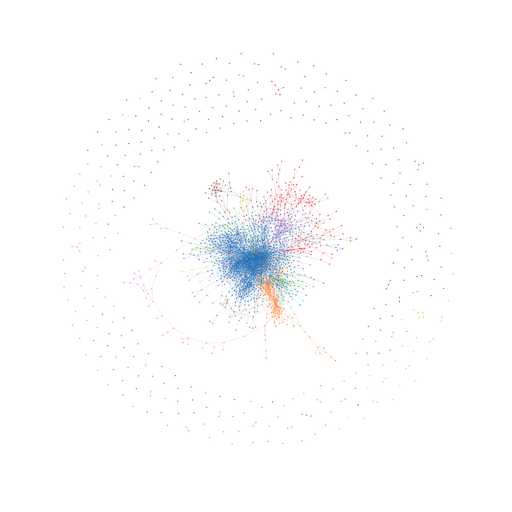}}
\end{subfigure}
\begin{subfigure}[2000 ($\langle k \rangle = 7.68$ and $n=4337$))]{
\includegraphics[width=0.31\linewidth]{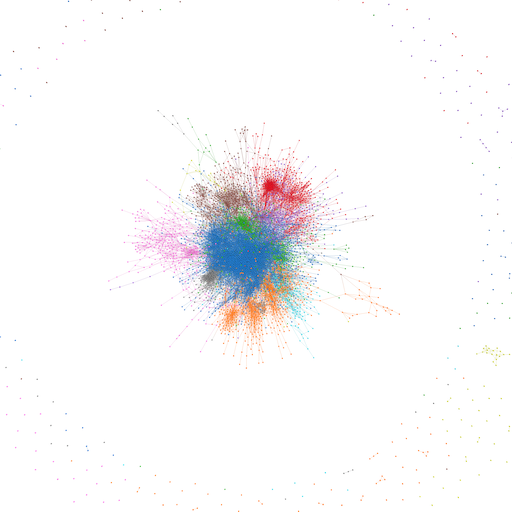}}
\end{subfigure}

\begin{subfigure}[2010 ($\langle k \rangle = 8.65$ and $n=6892$))]{
\includegraphics[width=0.31\linewidth]{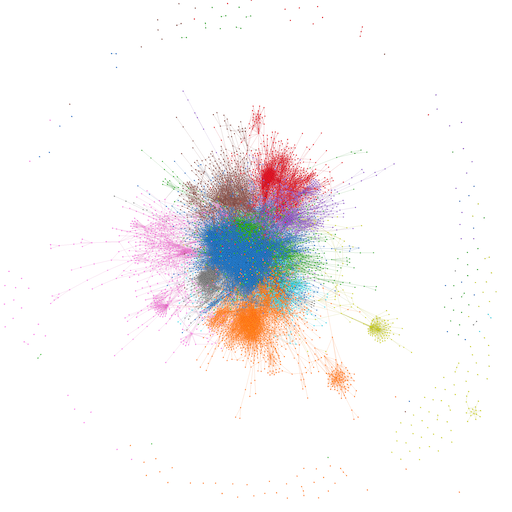}}
\end{subfigure}
\begin{subfigure}[2018 ($\langle k \rangle = 9.74$ and $n=8082$)]{
\includegraphics[width=0.31\linewidth]{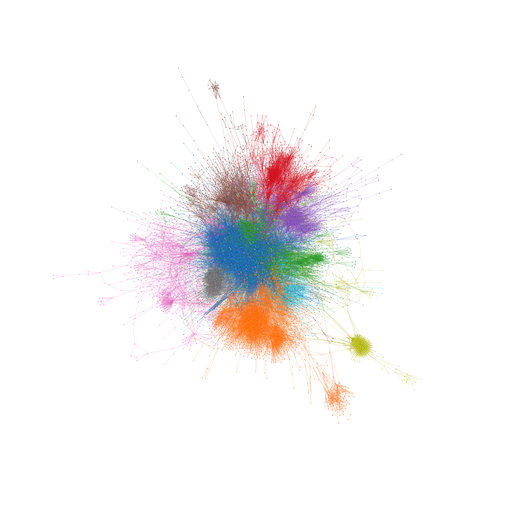}}
\end{subfigure}
\includegraphics[width=0.2\linewidth]{Legenda.pdf}
\caption{The citation network visualized along time. The average degree ($\langle k \rangle$) was calculated by considering the undirected version of the networks, and $n$ is the number of nodes.}
\label{fig:temp}
\end{figure*}

The time series in the earlier group are, as could be expected, more sparse, corresponding to the onset of several of the identified areas. The main area A, more directly related to bipolar transistors, has a peak around 1990, decreasing gradually thereafter. This could be interpreted as an indication of the come of age of bipolar technology around that time. A similar trend can be seen in the case of area D, which reaches a peak around 2000.  Also interesting is the increase, followed by a relative stabilization, observed for area B.
The above discussed temporal trends can be complemented by taking into account the changes in the topology of the respective overall network over time, which is illustrated in Figure~\ref{fig:temp} (a-h).

\begin{figure*}[!htpb]
  \centering
     \includegraphics[width=0.79\linewidth]{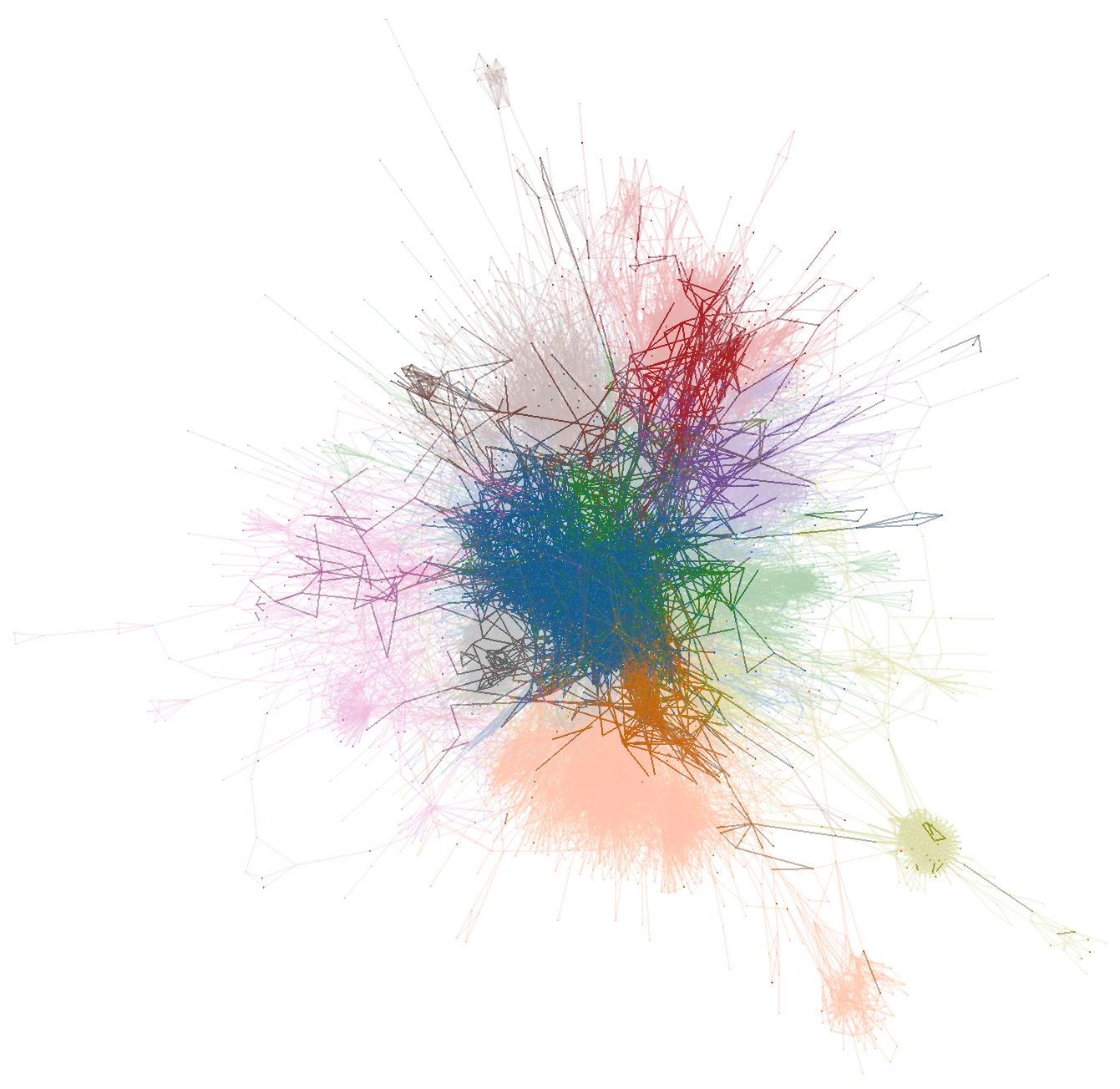}
     \includegraphics[width=0.2\linewidth]{Legenda.pdf}
   \caption{In order to try to better understand the peak of citations of the BJT area occurring around 1990, we obtained a partial network considering the nodes and interconnections along the period from 1985 to 1995, which is shown in darker colors, overlain onto the complete network (diluted colors).  Observe that the identified areas of analog electronics, voltage regulation, processing, and radiation control are resulted markedly pronounced in the partial network. This visualization
   is complemented by absolute and relative indications of the number of nodes and interconnections provided in Table~\ref{tab:size_highlight}.}
  \label{fig:highlight}
\end{figure*}

In order to better understand the possible reasons behind the peak observed
in Figure~\ref{fig:time_line}, we performed the following additional experiment.  We considered the entries and respective interconnections along the time period from 1985 to 1995, and obtained the respective partial network, which is depicted in Figure~\ref{fig:highlight} superimposed onto the more complete network, the latter being shown in diluted colors.  

This figure indicates that some areas, such as analog electronics as well as reference voltage areas (in red), seem to be particularly dense when compared to the other areas.  Table~\ref{tab:size_highlight} provide the total and relative number of nodes and connections accounted by the partial network (i.e.~extending from 1985 to 1995).  
We have from Figure~\ref{fig:highlight} and Table~\ref{tab:size_highlight} that the following identified areas are particularly pronounced, considering the adopted database, along the period from 1985 to 1995: (i) analog electronic; (ii)voltage regulation; (iii) processing; and (iv) radiation control  

Therefore, we hypothesize that the peak in the identified BJT area seems to be related to the coming of age of its relationship with analog electronics and other neighboring areas.

\begin{table*}[!htbp]
\scriptsize
\begin{tabular}{|c|c|c|c|c|c|}
\hline
\multirow{2}{*}{\textbf{Group name}} &  \multicolumn{3}{c|}{\textbf{Size}} & \multicolumn{2}{c|}{\textbf{Averange degree}} \\ 
\cline{2-6} 
 & \textbf{Entire network} & \textbf{From 1985 to 1995} & \textbf{proportion (\%)} 
 & \textbf{Entire network} & \textbf{From 1985 to 1995}  \\ \hline
A - Bipolar transistor & 2270 & 923 & 44.66 & 10.97 & 11.62\\
B - Power electronics & 1407 & 211 & 15.00 & 9.82 & 8.94 \\
C - Field Effect & 848 & 202 & 23.82 & 8.86 & 8.96 \\
D - Analog Electronics & 769 & 184 & 23.93 & 9.73 & 10.70 \\
E - Reference voltage & 644 & 91 & 14.13 & 10.07 & 10.89\\
F - High frequency & 629 & 128 & 20.35 & 6.95 & 7.39\\
G - Processing & 601 & 178 &  29.62  & 5.07 &4.93\\
H - Radiation control & 382 & 92 & 24.08 & 15.94 & 19.29  \\
I - Digital control & 269 & 36 & 13.38 & 11.08 & 7.83\\
J - Anti-statics & 263 & 55 & 20.91 &  7.42 & 8.32\\
\hline
\end{tabular}
\caption{Absolute and relative number of nodes in the partial (from 1985 to 1995) and complete networks.\label{tab:size_highlight}}
\end{table*}

The steady growth and consolidation of area A can again be noticed along the period extending from 1950 to 1990.  The other areas tend to follow the development of area A, though relatively few citations can be observed in the networks corresponding to these periods (many isolated nodes).
A steady increase in connectivity between these areas can be appreciated after 1990.  Remarkably, the relative distribution of the identified areas remained almost unaltered over the years, indicating that the interrelationship between those areas tended to preserve its pattern.

The identified area I, which is mostly constituted by patents (62.83\%) resulted particularly dense in connectivity, which was found to be related to the fact that six entries in that community receive citations for many of the other entries in that area.  In striking contrast, the other identified area incorporating many patents, namely community G, is particularly sparse.
  
\begin{figure}[!htpb]
  \centering
     \includegraphics[width=0.99\linewidth]{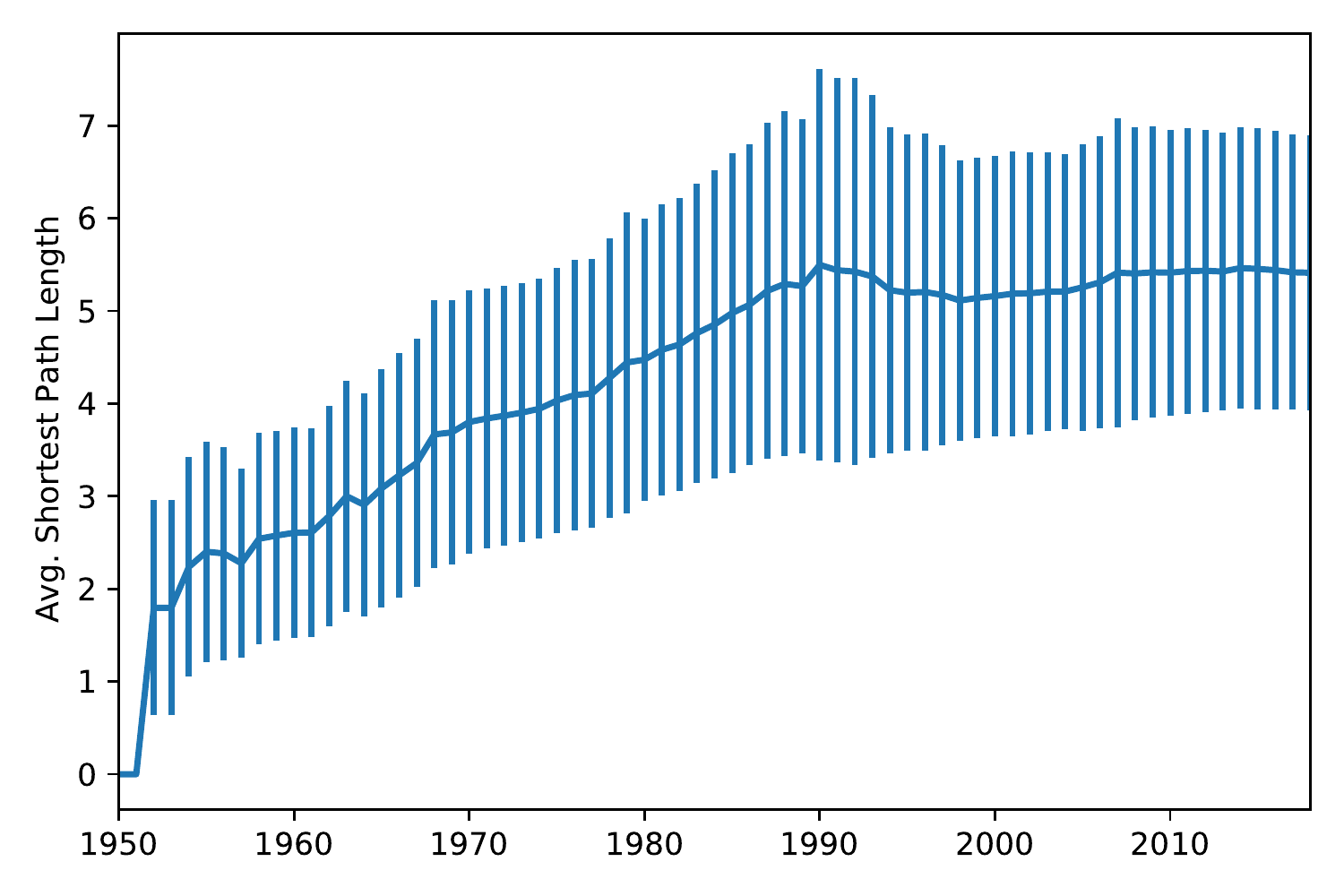}
   \caption{Mean shortest path lengths, obtained for the undirected version of the network. The vertical bars represent the standard deviations.}
  \label{fig:mean_shortest}
\end{figure}

Additional information about the interrelationship between the identified areas can be provided by the mean and standard deviation of the shortest path in the respective network, as depicted in Figure~\ref{fig:mean_shortest}. The mean shortest path can be observed to grow steadily from 1960 to 1990, stabilizing thereafter, corroborating the previous discussion.  We also observe an increase of shortest path length deviation around 1990, implying that at this time the network is not only particularly intensely connected, but these connections also exhibit an elevated degree of heterogeneity, as expressed by the respective standard deviation of the node degree. 

The noticeable stabilization of the network in Figure~\ref{fig:mean_shortest} from around 2000 suggests that most of the possible applications of electronics were by this time well covered.

\section{\label{sec:conc} Conclusions}

The modern world is has been, to a considerable extent, influenced, or even shaped by electronics.  The establishment of modern electronics was largely driven by the development of bipolar junction technology, which leads to widespread commercial applications of the respectively developed transistor. In the present work, we applied concepts and methods from network science, as well as science of science, to derive a tentative analysis of how several related areas started and evolved along time.

Several interesting results have been reported and discussed, which need to be understood in the context of the adopted dataset and methodology, as well as parametric configurations. In this restricted and tentative context, we observed a good thematic coherence between the entries obtained in each of the 10 identified areas, which were assigned respective tentative labels. These communities were then characterized in terms of several topological measurements including in- and out-degree, betweenness centrality, and clustering coefficient. We observed a surprising homogeneity in in- and out-degree obtained for each of the 10 communities. A more heterogeneous distribution was observed for the betweenness centrality, suggesting distinct patterns of interconnectivity among the communities.

The study of the identified areas along time revealed the onset and coming of age of the several identified areas, with the communities related to high frequency and digital control presenting a more recent pattern of growth.  A peak of publications was observed around 1990 for the reference area of bipolar junction technology.  The time evolution of the areas was also considered from the perspective of graph visualization along with different time instants, revealing surprising stability of the pattern of interconnections among the identified areas.

Several future developments are possible, including the extension of the analysis to other areas of electronics (e.g. integrated circuits, personal computers, telecommunications), or even other fields such as the Internet, artificial intelligence, among many other possibilities.

\section*{Acknowledgments}
Alexandre Benatti thanks Coordenação de Aperfeiçoamento de Pessoal de N\'ivel Superior - Brasil (CAPES) - Finance Code 001. Henrique F. de Arruda acknowledges FAPESP for sponsorship (grant no. 2018/10489-0 and no. 2019/16223-5). Luciano da F. Costa thanks CNPq (grant no. 307085/2018-0) and NAP-PRP-USP for sponsorship. This work has been supported also by FAPESP grants 2015/22308-2. Research carried out using the computational resources of the Center for Mathematical Sciences Applied to Industry (CeMEAI) funded by FAPESP (grant 2013/07375-0).

\bibliographystyle{ieeetr}
\bibliography{references}

\end{document}